\documentclass[preprint,12pt]{elsarticle}

\usepackage[T1]{fontenc}
\usepackage[latin9]{inputenc}
\usepackage{graphicx}
\usepackage{comment}
\usepackage[dvipsnames,usenames]{color}
\usepackage{draftwatermark}
\SetWatermarkText{Preprint}
\SetWatermarkScale{5}

\usepackage{background}

\backgroundsetup{
	angle=90,
	opacity=1,
	scale=1,
	color=black,
	nodeanchor=east,
	position={current page.west},
	contents={{Accepted to be published in \emph{Information Fusion}}},
	vshift=-20pt
}

\begin{document}
	
\begin{frontmatter}
	\title{Radar networks: A review of features and challenges}
	
	\author[a,b]{S.~Hamed~Javadi\corref{cor1}}
	\ead{h.javadi@ugent.be}
	\author[c]{Alfonso~Farina,~\emph{Fellow~of~EURASIP}}
	\ead{alfonso.farina@outlook.it}
	\cortext[cor1]{Corresponding author.}
	\address[a]{Department of Environment, Ghent University, Ghent, Belgium}
	\address[b]{Department of Electrical Engineering, University of Bojnord, Bojnord, Iran}
	\address[c]{Selex ES (now retired), Rome, Italy}

\begin{abstract}
Networks of multiple radars are typically used for improving the coverage and tracking accuracy. Recently, such networks have facilitated deployment of commercial radars for civilian applications such as healthcare, gesture recognition, home security, and autonomous automobiles. They exploit advanced signal processing techniques together with efficient data fusion methods in order to yield high performance of event detection and tracking. This paper reviews outstanding features of radar networks, their challenges, and their state-of-the-art solutions from the perspective of signal processing. Each discussed subject can be evolved as a hot research topic.
\end{abstract}

\begin{keyword}
Data fusion \sep detection \sep estimation \sep radar network \sep registration error \sep sensor management \sep signal processing \sep target tracking \sep wireless sensor network.
\end{keyword}

\end{frontmatter}

\section{Introduction}

Radars are mostly famous for their expansive use in  defense, air traffic control, weather monitoring and prediction, maritime control and aerial industries \cite{farina1986}. Recently, the development of commercial and low-cost radars,
such as those presented by \cite{Mencia-Oliva,Rodenbeck2005}, have
made it possible for many civilian applications to take advantage
of them. Though the commercial radars do not have the high-performance functionality and reliability of the military-class ones, advances in RF/microwave technologies and manufacturing have fit them for being used in even some defense applications \cite{lorell}.

\begin{figure}[ht]
	\begin{centering}
		\includegraphics[width=3in]{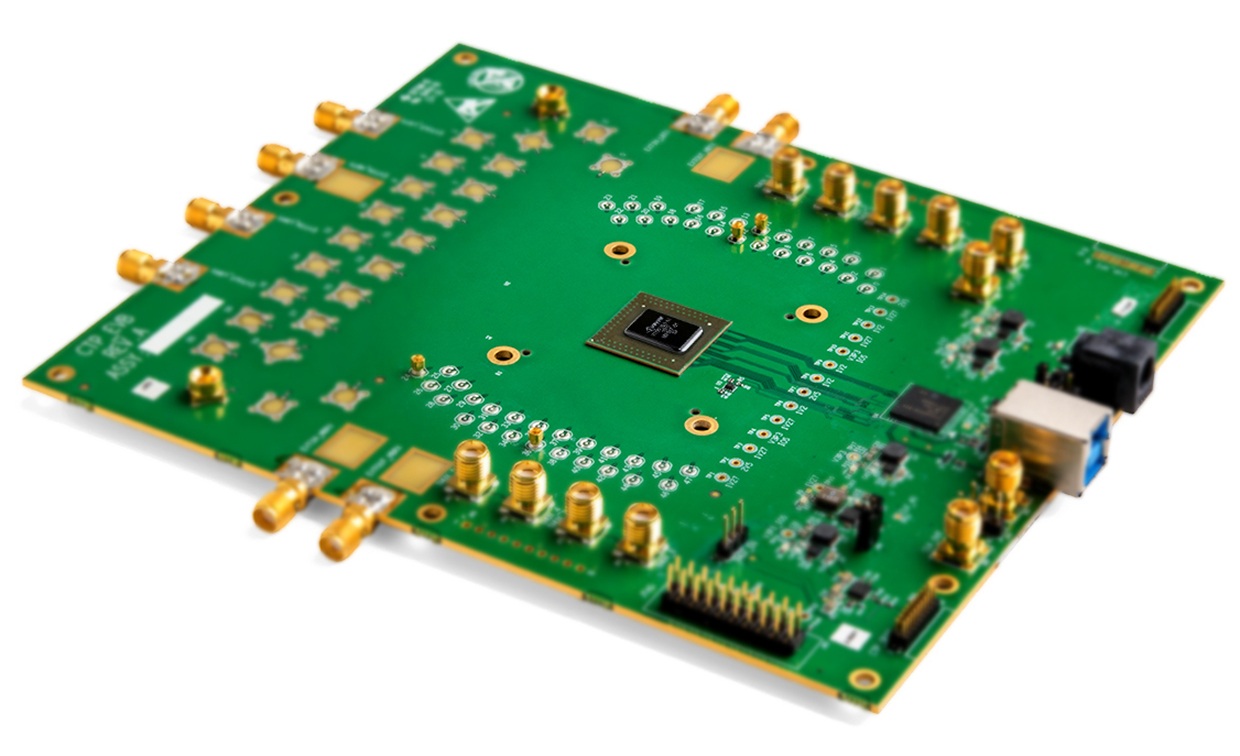}
		\par\end{centering}
	\caption{A commercial radar with 72 transceivers designed for smart home applications \cite{moore}. \label{fig:vayyar}}
	
\end{figure}

The civilian applications of radars are expanding. For example, while implementing surveillance cameras for
health-care induces privacy concerns, using radar sensors instead together
with advanced signal processing methods, such as the fall detection
method proposed in \cite{Garripoli,Gurbuz2019}, alleviate the issue \cite{Postolache}. Assisted living (AL) using radars consists of human activity recognition by signal processing and its classification by exploiting learning methods \cite{Kernec2019,Gurbuz2019}. A commercial radar for smart home applications designed by Vayyar has been shown in Fig. \ref{fig:vayyar} \cite{moore} with embedded on-board antennas.

\begin{figure*}[ht]
\begin{centering}
\includegraphics[width=4.5in]{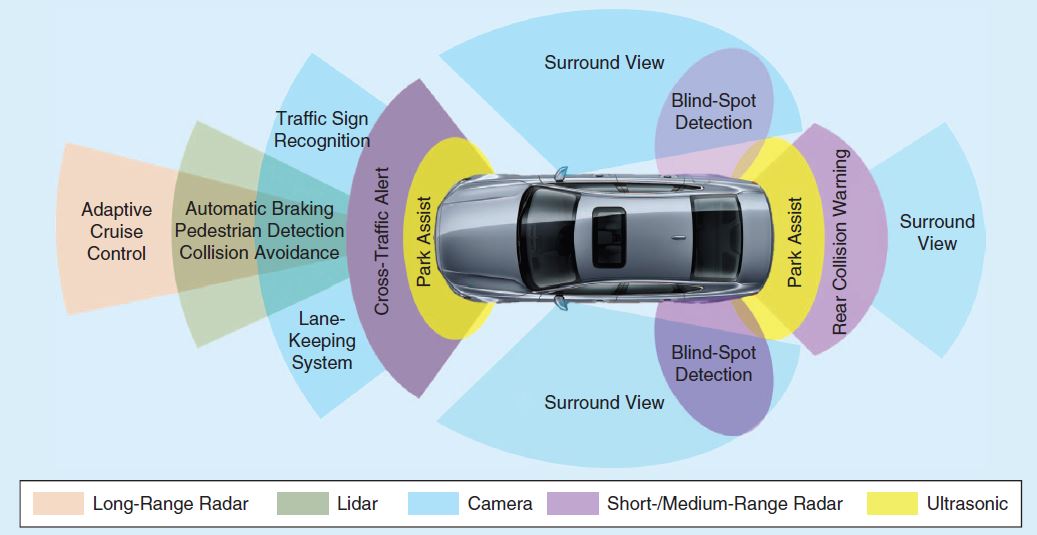}
\par\end{centering}
\caption{The application of radar networks in autonomous cars \cite{Kukkala2018}. \label{fig:The-application-of}}

\end{figure*}

Radars \textendash{} together with lidars \textendash{} have also
found applications in the pervasive autonomous vehicles \cite{Folster2005} (Fig. \ref{fig:The-application-of} \cite{Kukkala2018}).
Ward and Folkesson in \cite{Ward} have used radars for accurately localizing cars when they are out of visibility of GPS satellites. Wax in \cite{Wax} has discussed using radars for
detecting obstacles by autonomous vehicles. In addition, the Congress
of the United States had mandated that at least one third of military vehicles
must be autonomous by 2015 \cite{Wax}. 

\begin{figure}[ht]
\begin{centering}
\includegraphics[width=3.3in]{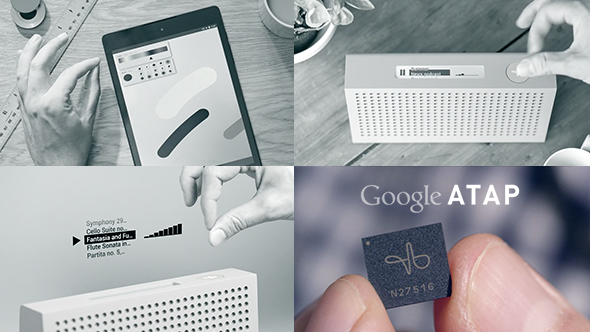}
\par\end{centering}
\caption{Gesture recognition using radars \cite{soli}. \label{fig:Gesture-recognition-using}}
\end{figure}

Another interesting application of radars is non-contact gesture
recognition \cite{Fan2016,Kim2016,Townley} (Fig. \ref{fig:Gesture-recognition-using} \cite{soli}).
Fan et. al. in \cite{Fan2016} have developed a remote
computer mouse by recognizing human gestures using a network of short-range continuous-wave (CW) Doppler radars. Kim and Toomajian in \cite{Kim2016} have proposed a method
to recognize hand gestures using micro-Doppler signatures measured
by Doppler radars. Moreover, designing a frequency modulated CW radar for recognizing gestures in mobile devices has been
explained by Townley et. al. in \cite{Townley,Allison2019} .

Commercial radars suffer from low coverage and deficit performance. Therefore, advanced signal processing methods are essential for efficiently exploiting them. In addition, multiple radar sensors are required in order to improve
both coverage and performance. Hence, efficient data fusion techniques should be employed as well.

\begin{figure}[ht]
	\begin{centering}
		\includegraphics[width=3.3in]{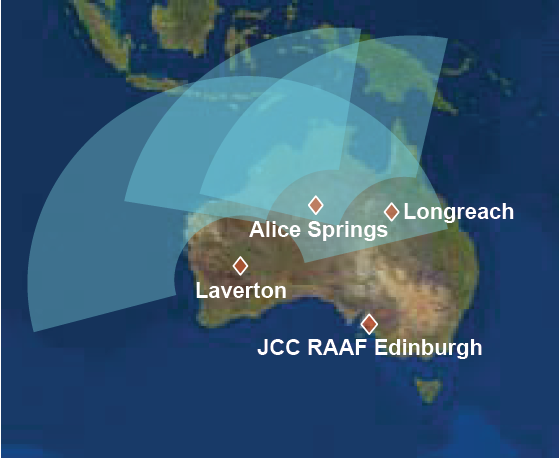}
		\par\end{centering}
	\caption{Jindalee operational radar network (JORN) for surveillance of Australia's northern approaches \cite{Allison2019}. \label{fig:jorn}}
\end{figure}

Jindalee Operational Radar Netowrk (JORN) \cite{Allison2019} --- shown in Fig. \ref{fig:jorn} --- is an interesting example of expanding surveillance coverage. In JORN, three over-the-horizon radar (OTHR) systems are used in remote areas of Australia in order to provide cost-effective surveillance of defense and civilian activities. Radar networks are also the efficient solutions to improve coverage while maintaining spatial resolution in weather forecasting and emergency management \cite{Junyent2009,Junyent2010}.

This paper presents the most important challenges of radar networks
and their state-of-the-art solutions from the perspective of \emph{signal processing}. Radar networks can be considered as a special class
of wireless sensor networks (WSNs) whose main goals usually consist
of detection and tracking of desired targets. Therefore, the challenges
of radar networks are categorized as: 
\begin{itemize}
\item Deployment of radars;
\item Decentralized detection;
\item Multi-target tracking (MTT);
\item Registration error correction;
\item Inference-driven sensor management.
\end{itemize}
Note that network aspects of radar networks --- such as temporal synchronization \cite{Djenouri2016} and spectrum availability \cite{Yucek2009,Awin2019} --- while are of practical importance, are not covered in this survey since there are currently many related works.

The paper is organized as follows. Deployment of radars is discussed
in Sec. \ref{sec:Deployment-of-radars}. Sec. \ref{sec:Decentralized-detection}
reviews problems related to decentralized detection over radar networks.
Tracking targets and relevant methods are discussed in Sec. \ref{sec:target.tracking}. Sec. \ref{sec:data.fusion} presents outstanding fusion
schemes with the methods to correct their registration errors explained in Sec. \ref{sec.registration.error}. The sensor management algorithms are reviewed in Sec. \ref{sec:Inference-driven-sensor-manageme} and finally the paper is concluded in Sec. \ref{sec:Conclusion}.

\section{Deployment of radars\label{sec:Deployment-of-radars}}

Generally, the deployment of sensors affects the overall performance
of network. Javadi in \cite{Javadi2016} has shown that a dense network
of cheap but available sensors can perform the same as a sparse network
of accurate but expensive sensors. The effects of network density
are also studied in \cite{Chamberland}. 

\begin{figure}[ht]
	\begin{centering}
		\includegraphics[width=3in]{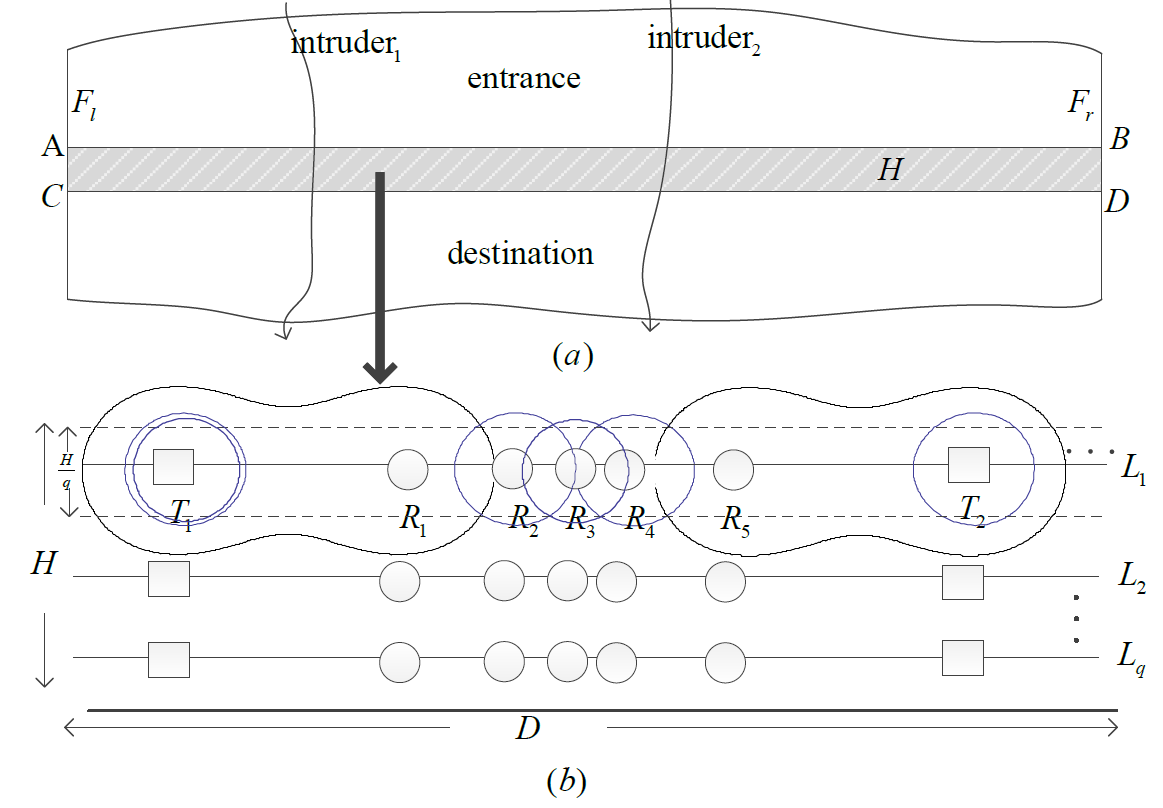}
		\par\end{centering}
	\caption{(a) Barrier coverage using a radar network \cite{Wang2016}. (b) The goal is to minimize the total cost of transmitters ($T_i$) and receivers ($R_i$). \label{fig:barrier}}
	
\end{figure}

One application of radar networks is in surveillance of barriers, such as borders, where the barrier coverage is crucial. Placement of bistatic radars
for barrier coverage has been investigated in \cite{Wang2016} as
an optimization problem where a minimum-cost placement strategy has been proposed
for full barrier coverage. It has been assumed that --- as shown in Fig. \ref{fig:barrier} \cite{Wang2016} --- the radar field covers the barrier breadth $H$ and $q$ deployment lines parallel to the long side of the sensor field are considered for placement of transmitters ($T_i$) and receivers ($R_i$).

One of the most popular sensor deployment methods is the Lloyd algorithm
\cite{Jorge2005,LloydWiki} whose goal is to minimize distortion in homogeneous
WSNs. Each sensor has a sensing range within which it can detect events
occurrence. However, the sensing capability of sensors decreases vs.
distance. Here, distortion is used for modeling the sensing inaccuracy.

The drawback of Lloyd algorithm is considering just the sensing range (coverage
aspect) while ignoring the connectivity (communication aspect). Guo and
Jafarkhani in \cite{Guo2016} have attempted to alleviate the connectivity
issue of Lloyd algorithm by considering both coverage and connectivity
simultaneously.

\section{Decentralized detection\label{sec:Decentralized-detection}}

Detection is a basic task of radar networks. In fact, target
detection and tracking are the main goals of radar networks. A comprehensive
tutorial on the topic could be found in \cite{Javadi2016}. In detection, the goal is to detect desired targets correctly while avoiding
any false alarm \cite{Maherin2015}. Detection may be carried out in either a centralized or decentralized manner \cite{Conte1983}.

In centralized detection, radar sensors send
raw observations to a fusion center (FC) where the final decision about
target presence is taken. However, sending raw observations from radar
sensors to FC imposes a large communication burden. Therefore,
radar sensors are usually set up to process their observations locally
and send a compressed data to FC (decentralized detection). Less
data transmission results in energy saving as well
since low-power radar sensors consume most of their energy during data wireless
communication. Energy consumption is crucial specially when the radar
sensors are powered by batteries. The less data is sent, the more
energy is saved in the cost of performance loss.

Designing an optimum decentralized detector network
consists of designing optimal local detectors and designing an optimal
fusion rule at FC. The problem is well-known to be intractable
in general \cite{Tsitsiklis1993} even in the simplest case of a two-sensor
network \cite{Tenney1981}. Rational assumptions such as statistical
independence conditioned on each hypothesis (target presence and the
null hypotheses) make the problem tractable. If the statistical information
of the target is available, the likelihood ratio test (LRT) yields the
optimal detection performance \cite{KayJan.1998} in radar sensors.
In the absence of any statistical information, radar sensors may use
a quantization method --- such as the hyper-plane quantization
\cite{Fang2009} --- to compress their observations. Then,
an appropriate fusion rule, such as counting rule \cite{Niu2006},
weighted decision fusion (WDF) \cite{Javadi2015}, generalized LRT
(GLRT) or the Rao test \cite{KayJan.1998}, may be implemented at
FC in order to infer the final decision.

\section{Target tracking \label{sec:target.tracking}}

Tracking multiple targets is a challenging problem since it needs implementing statistical filters with cumbersome computations. An overview of different target tracking methods developed during the last forty years has been presented by Farina et. al. in \cite{farina2017conf}. 

While the Kalman filter (KF) and its derivatives such as extended KF (EKF) and unscented KF (UKF) are common in estimation of
the positions of targets, an appropriate additional filter is crucial
for data association, i.e. discriminating targets from the clutter as well as from each other.
The simplest approach is the nearest neighbor standard filter (NNSF)
\cite{Blackman} which updates the current prediction with the nearest
measurement to it and ignores the other measurements. Here, ``nearest''
refers to the observation with the minimum normalized innovation. In fact,
NNSF implicitly assumes a low clutter rate and thus does not work well
in situations with high degree of clutter.

A more flexible alternate to NNSF is the probabilistic data association filter (PDAF) \cite{Bar-Shalom2009}. PDAF computes the probabilities of correct associations for the measurements and use them to compute the track by weighted averaging. Despite NNSF which chooses just one measurement, PDAF involves all measurements but each with a weight proportional to its probability of correct association. The probability of correct association is usually considered to be proportional to the normalized innovation of the association. A comparison between the performances of NNSF and PDAF in \cite{Bar-Shalom2009} shows that PDAF exhibits more robustness than NNSF in the presence of clutters.

A data association algorithm based on PDAF in a radar network tracking a single target has been proposed in \cite{Yan2020}. There, a centralized collaborative detection and tracking strategy is implemented in which each radar is allocated with a transmission power and a false alarm rate in order to meet the energy limitations of the network. To that end, an optimization problem with the Bayesian Cramer-Rao lower bound as the objective function is considered and a sub-optimal solution is obtained.

For tracking multiple targets, two extensions of NNSF are suboptimal nearest neighbor (SNN) \cite{Farina1985} and global nearest neighbor (GNN) \cite{Blackman1986}. SNN treats with data association like an assignment problem and relates each track to its nearest observation, starting from the least distance to a track. However, SNN solution is not guaranteed to be the optimal nearest neighbor scheme. Alternatively, the global optimal scheme is obtained in GNN using the Munkres algorithm \cite{Munkres1957}. Other sub-optimal, but faster, solutions to GNN are the auction algorithm \cite{Bertsekas1988} and the Jonker-Volgenant-Castanon (JVC) method \cite{Jonker1987}.

Joint PDAF (JPDAF) \cite{Bar-Shalom1995} is the extension of PDAF for tackling MTT in which the joint probabilities of association are computed and used for updating the predictions of the tracks. PDAF and JPDAF are useful for environments with considerable clutter. The performance of JPDAF, SNN, and GNN in tracking six closely-spaced F-18 fighters has been examined and compared in \cite{Leung1999}. The results showed that the algorithms exhibit similar performances while they use different approaches.

\begin{figure}[ht]
	\begin{centering}
		\includegraphics[width=2.5in]{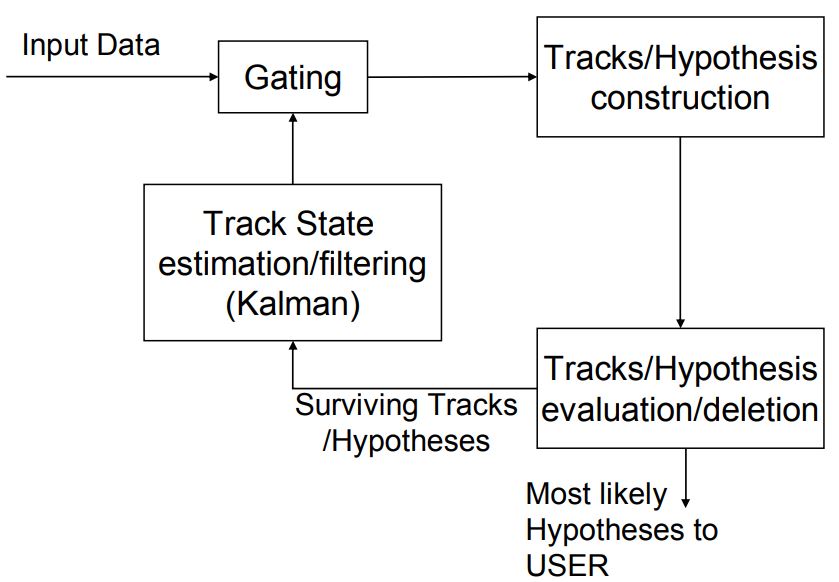}
		\par\end{centering}
	\caption{An overview of MHT \cite{Panta2004}. \label{fig:mht}}
	
\end{figure}

Another key algorithm of MTT is the multiple hypothesis tracking
(MHT) filter \cite{Reid1979, Bar-Shalom1995} which performs well in presence of clutter and high track uncertainties (e.g. maneuvering or crossing targets). MHT works as follows. A validation gate is computed based on the predicted observation and a new hypothetical track is established for each measurement inside the gate. The new tracks are treated independently and tracks with low likelihoods are discarded in order to avoid exponentially increasing the number of the tracks. An overview of MHT has been shown in Fig. \ref{fig:mht} \cite{Panta2004}. 

\begin{figure}[ht]
	\begin{centering}
		\includegraphics[width=3.5in]{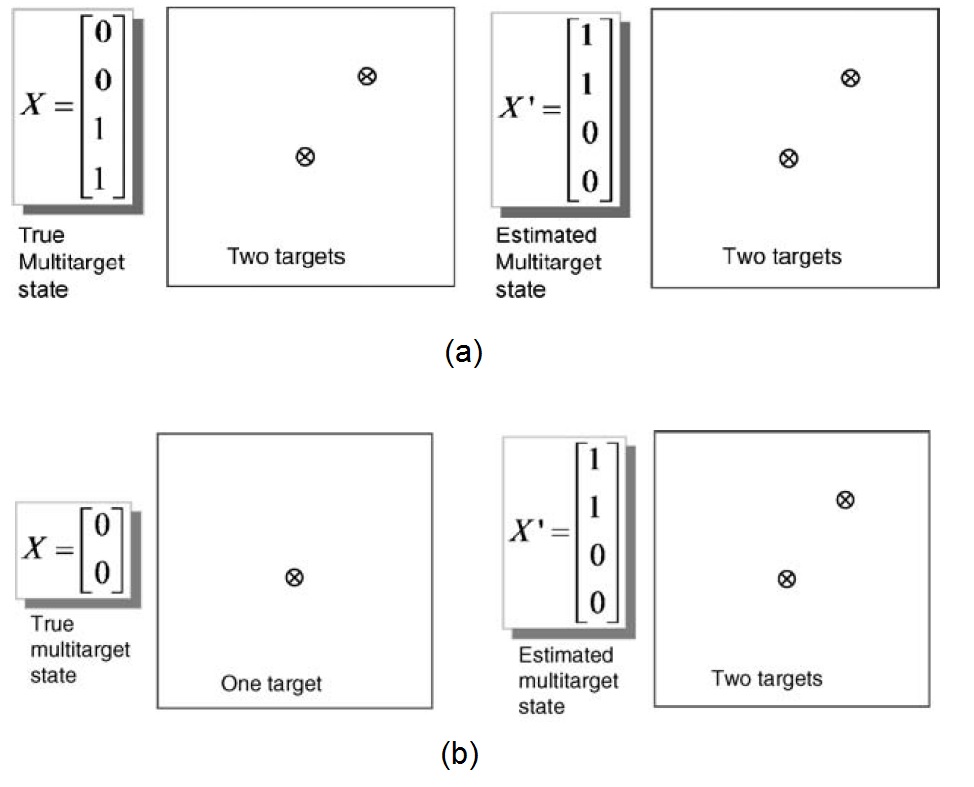}
		\par\end{centering}
	\caption{Estimation error when vectors are used as the multitarget state \cite{Mallick2012}. (a) Correct estimation with $\|X-X'\|=2$. (b) The true and estimated number of targets are not the same. \label{fig:est.error}}
	
\end{figure}

Calculating the multitarget estimation error becomes inconvenient when vectors are used for representing multitarget state --- as in the above discussed approaches. For example, see Fig. \ref{fig:est.error}-a \cite{Mallick2012} wherein an estimation error is mistakenly obtained while the estimation is correct. Another issue with using vectors is that they can not state miss distance when the estimated number of targets differs from the actual number of targets. As an example shown in Fig. \ref{fig:est.error}-b \cite{Mallick2012}, the true and estimated number of targets may not be equal. 

However, considering the states of targets as the elements of a set alleviates this problem well. For this purpose, random set theory \cite{Mahler,Mahler2014} --- also referred to as FInite Set STatistics (FISST) --- has been adopted in tackling the MTT problems.

In FISST, random finite sets (RFSs), instead of random vectors, are used in order to update both the
number and the states of the targets simultaneously. An RFS is a set with a random cardinality --- indicating the number of targets --- whose elements represent the state vectors of the targets. When resorting to FISST in an MTT problem, an appropriate metric should be defined for computing the distance between the set of the true states and the set of the estimated target states. Some of the metrics commonly used in MTT using RFSs include Hausdorff \cite{Chen2002Hausdorff}, Wasserstein \cite{Hoffman2004} --- also referred to as OMAT in \cite{Schuhmacher2008} ---, optimal sub-pattern assignment (OSPA), and generalized OSPA (GOSPA).
	
Hausdorff considers the largest of all the Euclidean distances from a point in one set to the closest point in the other set as the estimation error \cite{Chen2002Hausdorff}. Hence it is not sensitive to cardinality mismatches. Hoffman and Mahler alleviated the problem in \cite{Hoffman2004} by introducing a metric based on the Wasserstein distance which is originally a measure of similarity between probability distributions. The proposed Wasserstein distance has been improved in \cite{Ristic2011} where the optimal sub-pattern assignment (OSPA) metric is proposed in which the cardinality mismatches are penalized. A modification of OSPA is the generalized OSPA (GOSPA) \cite{Rahmathullah2017}.

While the traditional approaches solve the MTT problem by splitting it to multiple independent single-target tracking problems, the FISST-based method propagates all tracks simultaneously. To that end, a function known as the \emph{probability hypothesis density (PHD)} has a key role. In simple words, a PHD defines the intensity of targets in the state space and its integration over a region gives the expected number of targets in that region. A comparison between the performances of the PHD and MHT filters has been presented in \cite{Panta2004,Panta2009} via simulations. However, MHT is a flexible algorithm and its performance can be improved by modifying its parameters (e.g. in track deletion and merging as well as gating). Generally, there is not any valid comparison between the two algorithms.

While the PHD filter predicts and updates the PHD function
in time, another approach, the cardinalized PHD (CPHD) filter \cite{Mahler2007}, provides more robustness with respect to clutter and misdetections. The CPHD filter, which is more general than the PHD filter, updates the number of the targets in time in addition to updating just the PHD function and is preferable over PHD especially in large number of targets. Another FISST-based approach to MTT is the second order PHD (SO-PHD) \cite{Schlangen2018} in which the variance of the number of of targets is propagated in addition to its mean.

In another recent approach towards MTT, using the message passing algorithm has been proposed in \cite{Meyer2018}. Using this approach in terms of its sum-product algorithm (SPA), the conditional marginal posterior probability density functions (pdfs) (i.e. the pdf of each state given observations) can be computed; hence the prediction and update steps of tracking become tractable. It has been shown in \cite{Meyer2018} that this approach can deal with non-linear and non-Gaussian models as well. Moreover, the sum-product algorithm for data association (SPADA) proposed in \cite{Meyer2018} has been examined in several multi-object tracking applications \cite{Meyer2016,Meyer2017}. It is shown that SPADA performs well in cases with objects generating at most one measurement \cite{Meyer2019}.

\subsubsection*{Computational complexity}
The major part of computational complexity of MTT algorithms is imposed by data association. It has been shown in \cite{CONG19991} that the complexity of MHT is at least $\mathcal{O}\left\{\left(n_t|Z|\right)^2\right\}$ with $n_t$ and $|Z|$ denoting respectively the number of existing tracks and the number of measurements. This complexity is affected by track merging and measurement gating. On the other hand, it has been argued in \cite{Bar-Shalom2009} that the complexity of JPDAF is lower than that of MHT in terms of computation time and code memory size.

While the complexity of the CPHD recursion is very high in general, it is significantly reduced by considering the PHD function as the mixture of normally distributed states \cite{vo2007}. This implementation of CPHD is known as GM-CPHD and its complexity is $\mathcal{O}\left(|Z|^3\right)$. It has been claimed in \cite{vo2007} that GM-CPHD is simpler than JPDAF.

\subsubsection*{Sensor resolution}
A notable issue in MTT is sensor resolution which is defined as the ability to distinguish between two close targets \cite{Scala2002}. Most MTT algorithms ignore sensor resolution and assume it as perfect since considering it increases the computational complexity.

Sensor resolution is affected by the following factors \cite{Scala2002,Koch1997,Daum1994}:
\begin{itemize}
    \item Range resolution which depends on the length of the emitted pulse;
    \item Angular resolution which depends on the antenna beamwidth;
    \item Signal-to-noise ratio (SNR) for each target;
    \item The number of measurement samples;
    \item The algorithm implemented for tracking.
\end{itemize} 
On the other hand, if the probability of resolution is poor, the probability of data association will be heavily affected \cite{Daum1994}. Musicki et. al. in \cite{Musicki2008} have presented an algorithm for improving the resolution which yields a better tracking performance.

\section{Data fusion}\label{sec:data.fusion}
After one of the above MTT methods is adopted by a radar network
--- depending on the radar sensor's resources as
well as the environment circumstances --- , an efficient
fusion algorithm is required in order to obtain the overall tuned
tracks. Some of the most common fusion methods are listed as follows:
\begin{itemize}
\item \emph{Inverse covariance fusion (centralized tracking)} \cite{Manyika1994}: The radars send their \emph{observations} to FC where the track is obtained by usually exploiting KF or any of its derivatives such as EKF and UKF. In the KF filters, the process model has a key role. Accordingly and in order to make for the modeling errors, the modified strong tracking fusion (MSTF) has been proposed in \cite{Yang2016} in which the predicted estimation covariance is faded by a factor. The fading factor is obtained by equating the approximation of the actual innovation covariance with the theoretical one given by KF.

Implementation of centralized PHD filter for fusing is computationally expensive and needs several simplifying approximations \cite{Nannuru2018}. Alternatively, it has been proposed in \cite{Li2018} to preprocess the received observations and clustering them in sets of proxy and homologous measurements.

For cases with non-Gaussian process models, fusion of radars' observations by message passing algorithm has been presented in \cite{Meyer2018}.
\item \emph{Independent likelihood pool (ILP)} \cite{Manyika1994}: In ILP, radars send their likelihood ratios (LRs) to FC wherein the posterior
probability of the states is computed. The computed state may be
fedback to the radar sensors in order to compute their LRs more accurately.
ILP assumes the same prior information for all radars.
\item \emph{Covariance intersection (CI) fusion} \cite{Julier}: Known also as
``simple fusion'', ignores the correlation among the estimation
errors of radar sensors and thus significantly reduces the computational burden. CI is specially applicable in fusing radars' tracks when either the correlations between the state vectors of radars are unknown or there are restrictions on processing power or the memory size. It fuses the state vectors
of radars by weighting each according to their covariance matrix \cite{Deng2013}. Although consistent, i.e. more accurate than each local tracker \cite{Deng2013}, it yields a conservative track estimate \cite{Bakr2017}. Therefore, it has been modified in \cite{Bakr2018,Wu2018} by incorporating a partial prior knowledge regarding the unknown cross-correlations in order to improve the tracking performance.
\item Information graph \cite{chong1990}: Information graph is a solution to overcome the correlation existing in the data fused at FC by identifying redundant information. When several nodes report the same data, just one of them is new to FC while the others are redundant (double counting of information is also referred to as data incest). The common information of any two or more nodes could be found by identifying their common predecessors in the information graph. Then, the conditional dependencies from the data sets are removed in the Bayesian formulations \cite{chong1990}. This method is not applicable in networks with dynamic configurations.
\item \emph{Track-to-track fusion} \cite{Chang1997}: The track-to-track fusion
combines two state vectors while considering their correlation. It has been shown in \cite{Barshalom1986} that considering existing correlation between the two state vectors due to the common process noise improves the final estimation accuracy. This method is also referred to as the weighted covariance fusion (WCF) \cite{ChengWang2009} and has been extended to cases with more than two sensors in \cite{Chen2003}. It uses the state estimates together with the covariance matrix in order to obtain the fused state of the target.

The problem of fusing the tracks given by two different kinds of sensors while considering their cross-covariance, i.e. heterogeneous track-to-track fusion, has been considered in \cite{Mallick2019,Cormack2019}. In \cite{Mallick2019}, the tracks given by a passive infrared sensor and an active radar are fused using cubature Kalman filter (CKF).
\item \emph{Consensus-based fusion (distributed MTT \textendash{} DMTT)} \cite{Battistelli}:
In consensus-based fusion methods, all radars reach the same track
after several steps of information exchange with their neighbors.
FC is not needed anymore since fusion is carried out in a distributed
manner. Here, the goal of radar sensors is to locally update and fuse the
cardinality and the locations of targets such that the estimated values
are as close as possible to those given by the centralized
CPHD filter. Battistelli et. al. in \cite{Battistelli} have shown the considerable
tracking performance improvement reached after just one step of information
exchange.
\end{itemize}

\section{Registration error correction}\label{sec.registration.error}
Registration errors are referred to the errors due to any kind of uncertainty in local sensors such as asynchronous clocks, uncertainties in exact locations and orientations of radars and their measurement biases, and also the uncertainty in local settings of sensors. While some kinds of these errors, such as the error due to noise power uncertainty \cite{MOHAMMADI2019218} and uncalibrated sensors \cite{Ristic2012}, can be compensated locally, most of them must be corrected by FC. If not corrected, they may result in tracking errors and formation of multiple (ghosts) tracks on the same target \cite{Lin2004,Fortunati2011,Folster2005}.
	
When performing centralized tracking, the decoupled Kalman filter \cite{Shea2000} or the approach presented by \cite{Nabaa1999} may be adopted in order to compensate the bias errors of local sensors. These methods involve estimating cross-correlations and thus have a heavy computational burden. Another approach to estimate multi-radar biases in centralized tracking is resorting to FISST and using a PHD filter for estimation of the biases in a recursive manner \cite{Zhang2018}.

In track fusion, the more common approach to bias error correction estimates the bias errors (both offset and scale biases) dynamically by approximating the local measurements from their state estimates and then subtracting them from each other \cite{Lin2004}. This procedure gives a measurement of the biases which is independent of the states. Therefore, they can be estimated via either least squares estimation (LSE) or minimum mean square error (MMSE) estimation depending on whether the biases are modeled as unknown constants or random variables, respectively. This method has been extended in \cite{Lin2006} to asynchronous sensors. The case of asynchronous sensors has also been considered in \cite{Pu2018} where the registration problem has been tackled by non-linear optimization assuming a target with a nearly-constant-velocity model.

In the consensus-based fusion, which lacks any FC, a distributed registration method has been proposed in \cite{GaoBattistelli}. In this method, a cost function indicating the differences between local posteriors is minimized over registration errors.

\section{Inference-driven sensor management \label{sec:Inference-driven-sensor-manageme}}
\begin{figure}[ht]
	\begin{centering}
		\includegraphics[width=3.4in]{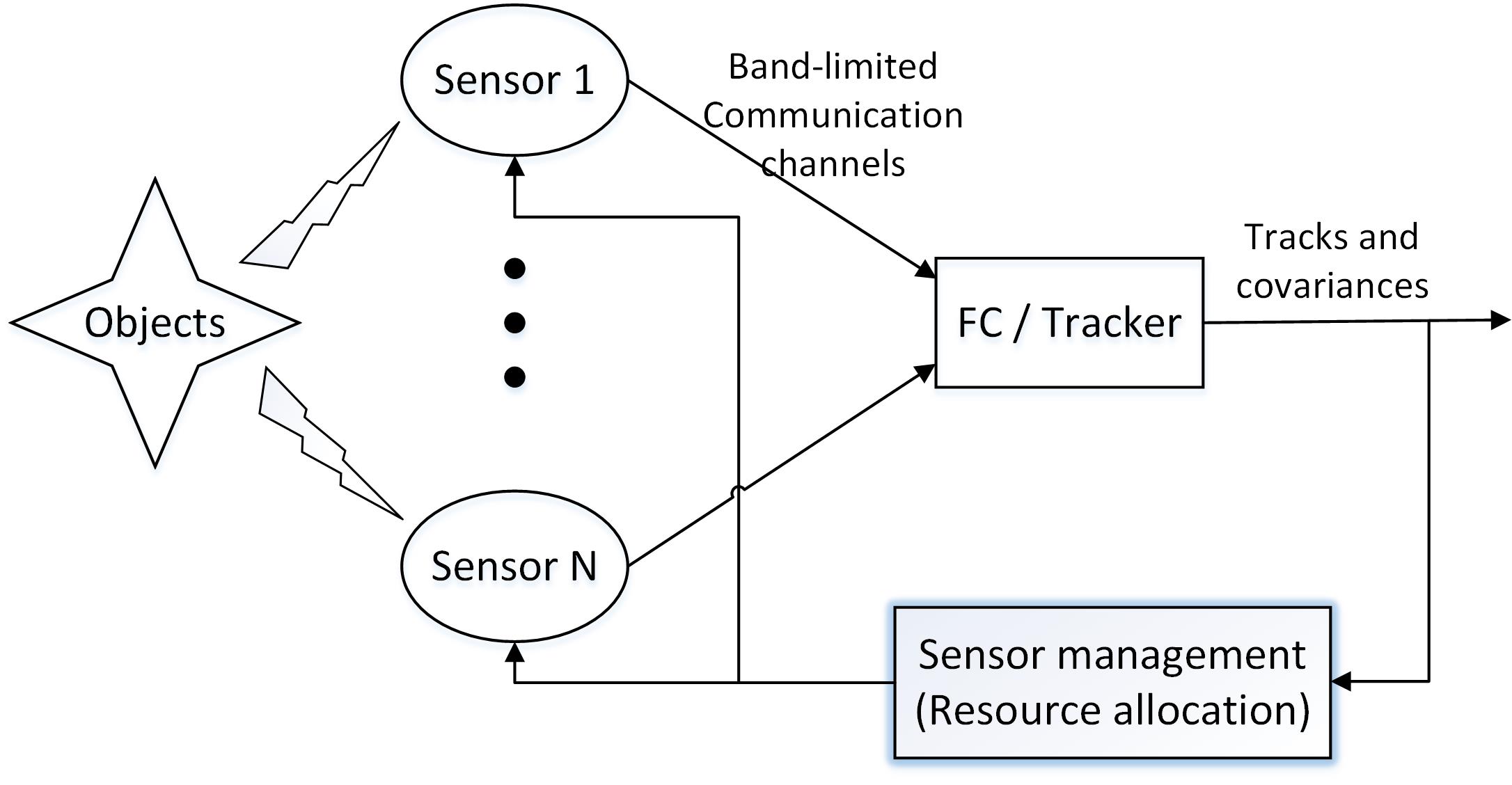}
		\par\end{centering}
	\caption{Managing resource-limited radar sensors and tasking a group of them to collect observations for statistical inference. \label{fig:sensor.management}}
	
\end{figure}

In radar networks, resources such as communication bandwidth and energy of radar motes (in case they are powered by batteries) are limited. Therefore, a sensor management mechanism would be useful for efficient use of network and prolonging its lifetime. Inference-driven sensor management refers to determining the optimal way of managing
the limited resources of the network, such as energy and bandwidth,
by assigning tasks to a determined group of sensors, as shown in Fig. \ref{fig:sensor.management}.

Sensor management is generally a non-convex optimization problem \cite{Joshi2009} with sub-optimal solutions reachable by convex relaxation as proposed by \cite{Joshi2009}. A framework based on multi-objective optimization has been proposed by Li et. al. in \cite{Li2019} with tracking accuracy and quantity budget (in terms of the number of the sensors selected) as the objective functions. In a distributed MTT scenario, a semi-definite-programming-based solution has been provided in \cite{Xie2018} where the feedback information in the tracking recursion is used for improving the worst-case tracking accuracy.

Other solutions to sensor management are mostly based on information
theoretic methods \cite{Yang2018}. Williams et. al. in \cite{Williams2007} assign
some sensors of the network as anchor nodes based on which a coarse
estimation of a desired source is obtained. Then, at each iteration, a set of (usually, a few) non-anchor sensors  are activated whose data maximizes the mutual information (MI) between the source location and the quantized sensors' measurements. After several iterations, the source is localized. 

\begin{figure}[ht]
	\begin{centering}
		\includegraphics[width=3.5in]{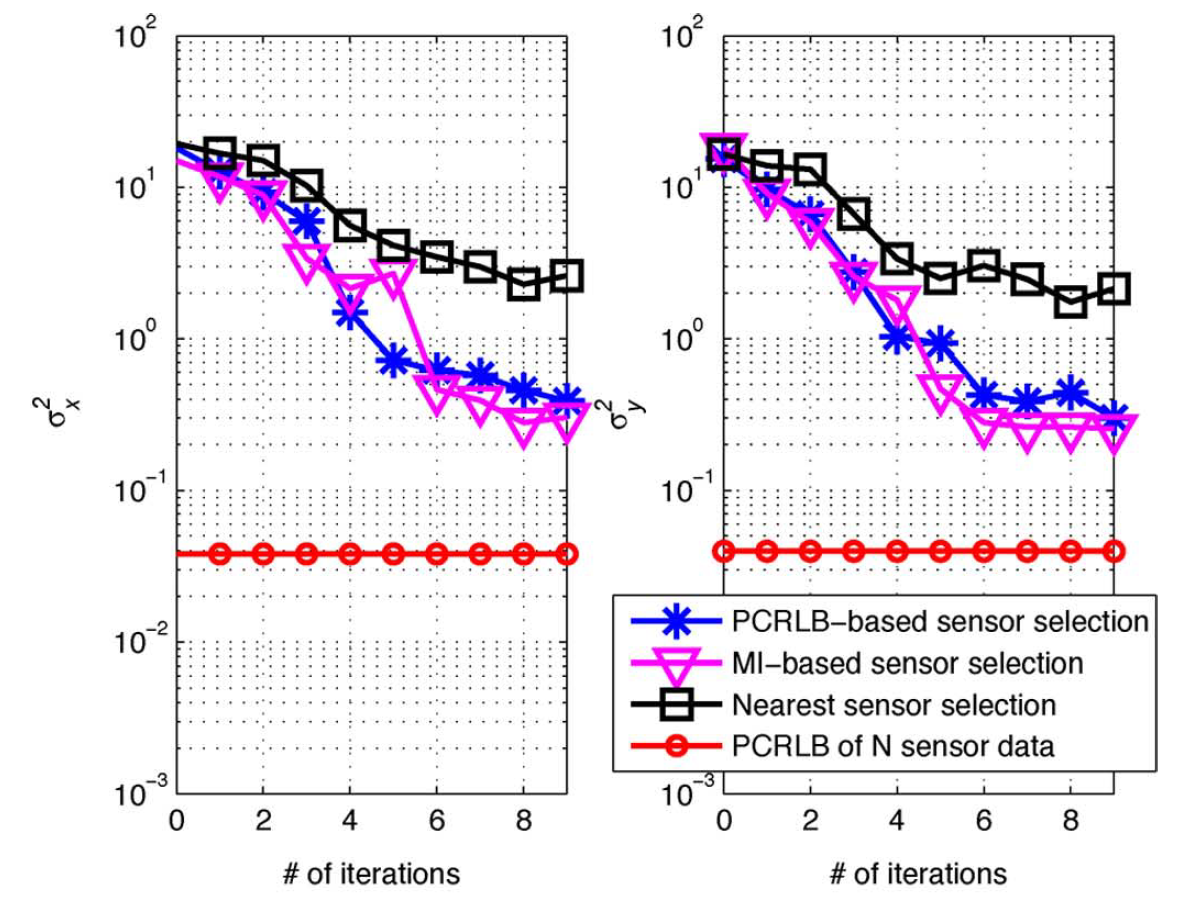}
		\par\end{centering}
	\caption{A comparison between MSE of different methods of sensor selection \cite{Masazade2010}. The isotropic signal model has been adopted for the source with energy 25000 and decay factor 2. The network consists of 361 sensors whose data is quantized using 3 bits. In each iteration of sensor selection, just one sensor is selected. \label{fig:sensor.selection}}	
\end{figure}

The MI-based method suffers from the perspective of complexity which grows
exponentially in the number of activated sensors in each iteration. Also, it can not be related to the final estimation performance. To alleviate these
issues, a sensor selection scheme based on the posterior Cramer-Rao
Lower Bound (PCRLB) has been proposed by \cite{Masazade2010,zuo2011}. In this method, a set of non-anchor sensors are activated that minimizes the PCRLB of the estimation error. The MI-based and PCRLB-based sensor management methods have been examined by \cite{Masazade2010} in a WSN with $361$ sensors deployed in a grid layout over a $100 \times 100m^2$ field. The iterative algorithms are initialized with $16$ anchor sensors and just one sensor is activated in each iteration. Fig. \ref{fig:sensor.selection} \cite{Masazade2010} shows the comparison between the MI-based and PCRLB-based methods with the nearest sensor selection method in term of MSE of $x$ and $y$ coordinates. In the nearest sensor selection method, sensors with the nearest measurements to the state track are selected.

In PCRLB, the Fisher information matrix (FIM) is obtained by taking expectation regarding both measurements and states. Therefore, PCRLB is an off-line measure independent of the state track. To take the state track into account, it has been proposed in \cite{zuo2011} to use the conditional PCRLB (C-PCRLB) in which FIM is conditioned on all past measurements. While the computational complexity of MI-based sensor selection rises exponentially with the number of activated sensors, it increases linearly in PCRLB and C-PCRLB-based methods. 

In realistic cases where sensor detection probability is less than unity and clutter exists in target tracking, the less-optimistic PCRLB formulations presented in \cite{Hernandez2006,Hernandez2017} may be used.

\section{Conclusion and future directions \label{sec:Conclusion}}
In this manuscript, the vital features and challenges of radar networks and their state-of-the-art solutions were reviewed from the perspective of signal processing. To that end, we categorized the existing broad literature into the following classes:
\begin{itemize}
\item Deployment of radars;
\item Decentralized detection;
\item Multi-target tracking (MTT);
\item Registration error correction;
\item Inference-driven sensor management.
\end{itemize}
Each class covers extensive topics that are currently being developed.
Evolving the solutions is crucial because of the rapidly expanding applications of radar networks. Any solution must meet the limitations of radar networks, especially limitations related to the communication burden and scalibility issues.

However, the topics related to radar networks are so extensive that it is not possible to cover them all in a review article. The subjects that were not covered by the current review include:
\begin{itemize}
    \item Selection of coordinate reference systems;
    \item Group tracking;
    \item Track initiation logics;
    \item Move stop move tracking \cite{Hernandez2011move};
    \item Multipath;
    \item Availability of software tools for tracking \cite{Crouse2017}.
\end{itemize}

The methods and algorithms related to radar networks may move forward by developing computational models and integrating artificial intelligence (AI) in data fusion. Their applicability in applications such as space surveillance can be considered as another future topic. 

To conclude, lots of information can be provided by a network of radars while they, hopefully, are at low cost due to the exploitation of technology of commercial products, and resistant against harsh environmental conditions. The accuracy of the information is improved by using advanced signal processing methods. It is hoped that the overall performance will even be improved more by integrating AI methods. The more accurate information facilitates decision making whose implementation in an autonomous manner is an open research field.

\bibliographystyle{elsarticle-num}
\bibliography{radar_networks}

\end{document}